\documentclass[11pt]{article}
\usepackage[margin=1in]{geometry}
\usepackage{graphicx}
\usepackage[normalem]{ulem}
\usepackage{tikz-cd}

\usepackage{adjustbox}

\usepackage{amsmath,amssymb}
\usepackage{hyperref}
\hypersetup{
colorlinks = true,
linkcolor = violet,
citecolor = violet,
urlcolor = violet
}

\newcommand{\R}{\mathbb{R}}

\newcommand{\xx}{\mathbf{x}}

\newcommand{\XX}{\mathbf{X}}

\title{Using Zigzag Persistent Homology to Detect Hopf Bifurcations in Dynamical Systems}

\author{
  Sarah Tymochko$^1$, Elizabeth Munch$^{1,2}$, Firas A. Khasawneh$^3$
}

\date{
$^1$ Dept. of Computational Mathematics, Science and Engineering\\
$^2$ Dept. of Mathematics\\
$^3$ Dept. of Mechanical Engineering \\
Michigan State University\\
\{tymochko,muncheli,khasawn3\}@egr.msu.edu
}

\begin{document}

\maketitle

\begin{abstract}
    Bifurcations in dynamical systems characterize qualitative changes in the system behavior.
Therefore, their detection is important because they can signal the transition from normal system operation to imminent failure.
While standard persistent homology has been used in this setting, it usually requires analyzing a collection of persistence diagrams, which in turn drives up the computational cost considerably.
Using zigzag persistence, we can capture topological changes in the state space of the dynamical system in only one persistence diagram.
Here we present Bifurcations using ZigZag (BuZZ), a one-step method to study and detect bifurcations using zigzag persistence.
The BuZZ method is successfully able to detect this type of behavior in two synthetic examples as well as an example dynamical system.
\end{abstract}

\section{Introduction}

Topological data analysis (TDA) is a field consisting of tools aimed at extracting shape in data.
Persistent homology, one of the most commonly used tools from TDA, has proven useful in the field of time series analysis.
Specifically, persistent homology has been shown to quantify features of a time series such as periodic and quasiperiodic behavior \cite{Perea2015,Sanderson2017,Tempelman2019,Maletic2016,Xu2018} or chaotic and periodic behavior \cite{Myers2019,Khasawneh2015}.
Existing applications in time series analysis include studying machining dynamics \cite{Khasawneh2017,Khasawneh2018a,Yesilli2019b,Khasawneh2015,Yesilli2019a,Khasawneh2018b,Khasawneh2014a}, gene expression \cite{Perea2015, Berwald2014}, financial data \cite{Gidea}, video data \cite{tralie2018quasi,tralie2016high}, and sleep-wake states \cite{Chung2019, Tymochko2020}.
These applications typically involve summarizing the underlying topological shape of each time series in a persistence diagram then using additional methods to analyze the resulting collection of persistence diagrams.
While these applications have been successful, the task of analyzing a collection of persistence diagrams can still be difficult.
Many methods have been created to convert persistence diagrams into a form amenable for machine learning \cite{Bubenik2015,Adams2017,Reininghaus2015,Perea2019}.
However, so many methods have been developed, it can be difficult to choose one appropriate for the task.
Additionally, the task of computing numerous persistence diagrams is computationally expensive.

Our method aims to circumvent these issues using zigzag persistence, a generalization of persistent homology that is capable of summarizing information from a sequence of point clouds in a single persistence diagram.
While less popular than standard persistent homology, zigzag persistence has been used in applications, including studying optical flow in computer generated videos \cite{Adams2019,Adams2020}, analyzing stacks of neuronal images \cite{Mata2015} and comparing different subsamples of the a dataset \cite{Tausz2011}.
However, to the best of our knowledge, it has not been used in the context of dynamical systems or time series analysis.
In this paper, we present Bifurcations using ZigZag (BuZZ), a one-step method to analyze bifurcations in dynamical systems using zigzag persistence.

\section{Materials and Methods}

Here we will present tools needed to build our method, including the time delay embedding, and an overview of the necessary topological tools.
Specifically, we briefly introduce homology, persistent homology and zigzag persistent homology.
However, we will not go into detail and instead direct the interested reader to \cite{Hatcher,Edelsbrunner2010,Carlsson2010} for more detail on homology, persistent homology, and zigzag homology, respectively.

\subsection{Homology and persistent homology}
\label{ssec:homology}

Homology is a tool from the field of algebraic topology that encodes information about shape in various dimensions.
Zero-dimensional homology studies connected components, 1-dimensional homology studies loops, and 2-dimensional homology studies voids.
Persistent homology is a method from TDA which studies the homology of a parameterized space.

For the purposes of this paper, we will focus on persistent homology applied to point cloud data.
Here, we need only assume a point cloud is a collection of points with a notion of distance, however in practice, this distance often arises from a point cloud in Euclidean space inheriting the ambient metric.
Given a collection of points, we will build connections between points based on their distance.
Specifically, we will build simplicial complexes, which are spaces built from different dimensional simplices.
A 0-simplex is a vertex, a 1-simplex is an edge, a 2-simplex is a triangle, and in general, a $p$-simplex is the convex hull of $p+1$ affinely independent vertices.
Abstractly, a $p$-simplex can be represented by the set of $p+1$ vertices it is built from.
So a simplicial complex, $\mathcal{K}$, is a family of sets that is closed under taking subsets.
That is, given a $p$-simplex, $\sigma$, in $\mathcal{K}$, then any simplex consisting of a subset of the vertices of size $0<k\leq p$, called a $k$-dimensional face of $\sigma$, is also in $\mathcal{K}$.

To create a simplicial complex from a point cloud, we use the Vietoris-Rips complex (sometimes just called the Rips complex).
Given a point cloud, $X$, and a distance value, $r$, the Vietoris-Rips complex, $\mathcal{K}_r$, consists of all simplices whose vertices have maximum pairwise distance at most $r$.
Taking a range of distance values, $r_0\leq r_1\leq r_2\leq \cdots r_n$ gives a set of simplicial complexes, $\{\mathcal{K}_{r_i}\}$.
Since the distance values are strictly increasing, we have a nested sequence of simplicial complexes
\begin{equation}\label{eqn:filtration}
\mathcal{K}_{r_0} \subseteq \mathcal{K}_{r_1} \subseteq \cdots \subseteq \mathcal{K}_{r_n}
\end{equation}
called a filtration.
Computing $p$-dimensional homology, $H_p(\mathcal{K})$, for each complex in the filtration gives a sequence of vector spaces and linear maps,
\begin{equation}
    H_p(\mathcal{K}_{r_0}) \to H_p(\mathcal{K}_{r_1}) \to \cdots \to H_p(\mathcal{K}_{r_n}).
\end{equation}
Persistent homology tracks homological features such as connected components and loops as you move through the filtration.
Specifically, it records at what distance value a feature first appears, and when a feature disappears or connects with another feature.
These distance values are called the ``birth'' and ``death'' times respectively.
These birth and death times are represented as a persistence diagram, which is a multiset of the birth death pairs $\{(b,d)\}$.

\subsection{Time delay embedding}
\label{ssec:timedelay}

One way to reconstruct the underlying dynamics given only a time series is through a time delay embedding.
Given a time series, $[x_1,\ldots,x_n]$, a choice of dimension $d$ and lag $\tau$, the delay embedding is the point cloud $\XX = \{\xx_i:= (x_i,x_{i+\tau}, \ldots, x_{i+(d-1)\tau}) \} \subset \R^d$.
Takens' theorem \cite{Takens1981} shows that given most choices of parameters, the embedding retains the same topological structure as the state space of the dynamical system and that this is in fact a true embedding in the mathematical sense.
In practice, not all parameter choices are optimal, so heuristics for making reasonable parameter choices have been developed \cite{Fraser1986,Kennel,Pecora2007,Chelidze2017,Pan2020,Kim1999}.

In existing methods combing TDA with time series analysis, most works analyze a collection of time series by embedding each one into a point cloud using the time delay embedding.
However, this can be computationally expensive and requires an analysis of the collection of computed persistence diagrams.
Instead, we will employ a generalized version of persistent homology to avoid these additional steps.

\subsection{Zigzag persistent homology}
\label{ssec:zigzag}

Zigzag persistence is a generalization of persistent homology that can study a collection of point clouds simultaneously.
In persistent homology, you have a nested collection of simplicial complexes as in Eqn.~\ref{eqn:filtration}.
However, for zigzag persistence, you can have a collection of simplicial complexes where the inclusions go in different directions.
Specifically, the input to zigzag persistence is a sequence of simplicial complexes with maps,
\begin{equation}
    \mathcal{K}_0 \leftrightarrow \mathcal{K}_1 \leftrightarrow \cdots \leftrightarrow \mathcal{K}_n
\end{equation}
where $\leftrightarrow$ is either an inclusion to the left or to the right.
While in general these inclusions can go in any direction in any order, for this paper we will focus on a specific setup for the zigzag based on a collection of point clouds.

\begin{figure}
    \centering
    \includegraphics[width=0.75\textwidth]{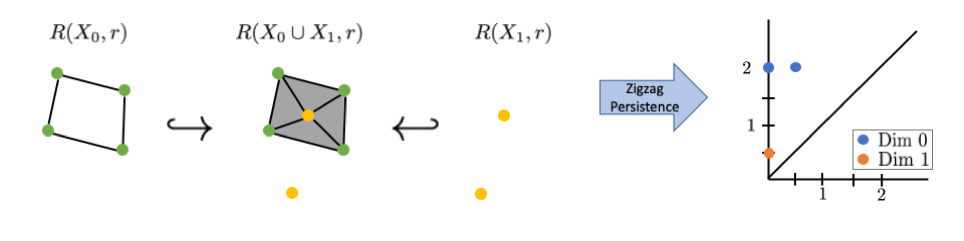}
    \caption{Small example of zigzag filtration with corresponding zigzag persistence diagram.}
    \label{fig:zz_example}
\end{figure}

Given an ordered collection of point clouds, $X_0,X_1,\ldots, X_n$, we can define a set of inclusions,
\begin{equation} \label{eqn:zz_pc}
    \begin{tikzcd}[column sep=tiny]
        X_0 \arrow[hookrightarrow,dr] &
         &
        X_1 \arrow[hookrightarrow,dl]\arrow[hookrightarrow,dr] &
         &
         \arrow[hookrightarrow,dl] X_2 &
         \cdots & X_{n-1}
         \arrow[hookrightarrow,dr] &
         &
        \arrow[hookrightarrow,dl] X_n \\
          &
         X_0 \cup X_1 &
          &
         X_1 \cup X_2 &
          & & &
         X_{n-1}\cup X_n. &
      \end{tikzcd}
\end{equation}
However, these are all still point clouds, which have uninteresting homology.
Thus, we can compute the Vietoris-Rips complex of each point cloud for a fixed radius, $r$.
This results in the diagram of inclusions of simplicial complexes
\begin{equation} \label{eqn:zz_rips_fixed}
    \begin{tikzcd}[column sep=-1.5em]
        R(X_0,r) \arrow[hookrightarrow,dr]
        &
        & R(X_1,r) \arrow[hookrightarrow,dl]\arrow[hookrightarrow,dr]
        &
        & R(X_2,r) \arrow[hookrightarrow,dl]
        & ~~~~~\cdots~~~~~
        & R(X_{n-1},r) \arrow[hookrightarrow,dr]
        &
        & R(X_n,r) \arrow[hookrightarrow,dl] \\
          & R(X_0 \cup X_1,r)
        &
        & R(X_1 \cup X_2,r)
        &
        &
        &
        & R(X_{n-1}\cup X_n,r).
        &
      \end{tikzcd}
\end{equation}
Computing the 1-dimensional homology of each complex in Eqn.~\ref{eqn:zz_rips_fixed} will result in a zigzag diagram of vector spaces and induced linear maps,
\begin{equation} \label{eqn:zz_rips_hom}
    \begin{tikzcd}[column sep=-1.5em]
        H_1( R(X_0,r) ) \arrow[dr]
        &
        & H_1( R(X_1,r) ) \arrow[dl]
        & ~~~~~\cdots~~~~~
        & H_1( R(X_{n-1},r) ) \arrow[dr]
        &
        & H_1( R(X_n,r) ) \arrow[dl] \\
          & H_1( R(X_0 \cup X_1,r) )
        &
        &
        &
        & H_1( R(X_{n-1}\cup X_n,r) ).
        &
      \end{tikzcd}
\end{equation}
Zigzag persistence tracks features that are homologically equivalent through this zigzag.
This means it records the range of the zigzag filtration where the same feature appears.
The zigzag persistence diagram records ``birth'' and ``death'' relating to location in the zigzag.
If a feature appears in $R(X_i,r),$ it is assigned birth time $i$, and if it appears at $R(X_i\cup X_{i+1},r),$ it is assigned birth time $i+0.5$.
Similarly, if a feature last appears in $R(X_j,r)$, it is assigned a death time $j+0.5$, while if it last appears in $R(X_j \cup X_{j+1},r)$, it is assigned a death time of $j+1$.
A small example is shown in Fig.~\ref{fig:zz_example}.
In this example, there is a 1-dimensional feature that appears in $R(X_0,r)$ and disappears going into $R(X_0 \cup X_1,r)$, thus it appears in the zigzag persistence diagram as the point $(0,0.5)$.
Additionally, there is one connected component that exists through the zigzag, corresponding to the 0-dimensional persistence point $(0,2)$.
By default, we assume all features die at the end of the zigzag, rather than having infinite points as in standard persistence.
There is one additional connected component that first appears in $R(X_0\cup X_1,r)$, corresponding to the 0-dimensional persistence point $(0.5,2).$

Note that we can easily generalize this idea to use a different radius for each Rips complex, $R(X_i,r_i)$.
For the unions we choose the maximum radius between the two individual point clouds, $R(X_i\cup X_{i+1}, \max\{r_i,r_{i+1}\})$, to ensure the inclusions hold.

\subsection{Bifurcations using ZigZag (BuZZ)}
\label{ssec:ourmethod}

\begin{figure}
    \centering
    \includegraphics[width=\textwidth]{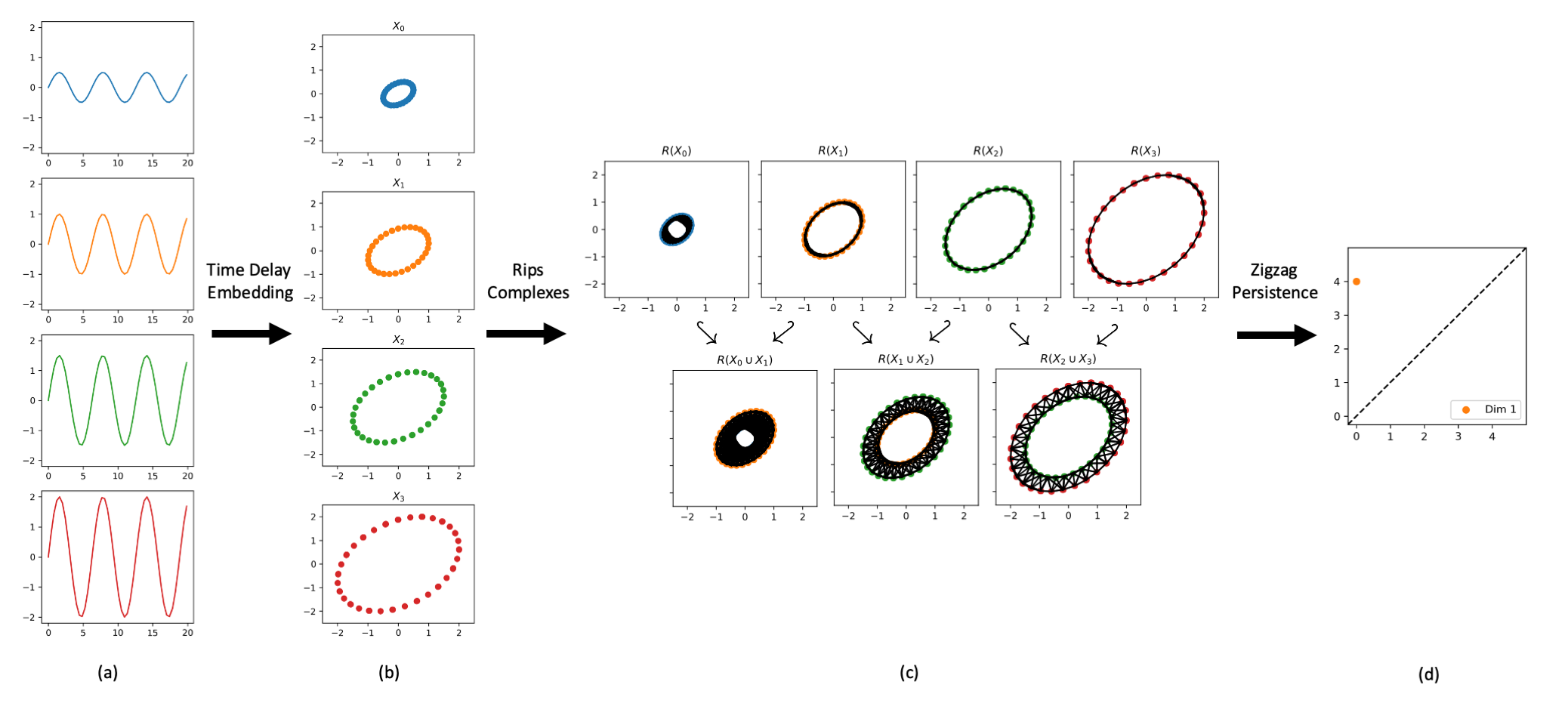}
    \caption{
    Outline of BuZZ method. The input time series is converted to an embedded point cloud via the time delay embedding. The Rips complexes are constructed for either a fixed $r$ or a choice of $r_i$ for each point cloud. Then, the zigzag persistence diagram is computed for the collection. }
    \label{fig:pipeline}
\end{figure}

We can now present our method, Bifurcations using ZigZag (BuZZ) for combining the above tools to detect changes in circular features in dynamical systems.
We will focus on Hopf bifurcations \cite{Guckenheimer1983}, which are seen when a fixed point loses stability and a limit cycle is introduced.
These types of bifurcations are particularly topological in nature, as the state space changes from a small cluster, to a circular structure, and sometimes reduces back to a cluster.

The necessary data for our method is a collection of time series for a varying input parameter value, as shown in Fig.~\ref{fig:pipeline}(a).
This particular example is a collection of time series given by $\{a\sin(t)\}$ for $a=0.5,1.0,1.5,2.0$ (going from top to bottom).
Each time series is then embedded using the time delay embedding (shown in Fig.~\ref{fig:pipeline}(b) using $d=2$ and $\tau=3$).
While in general, the delay could be varied for each time series, the embedding dimension needs to be fixed so each time series is embedded in the same space.
For the sake of interpretability and visualization, we will use a dimension of $d=2$ throughout this paper.
Sorting the resulting point clouds based on the input parameter value, the zigzag filtration can be formed from the collection of point clouds, as shown in Fig.~\ref{fig:pipeline}(c).
Lastly, computing zigzag persistence gives a persistence diagram, as shown in Fig.~\ref{fig:pipeline}(d), encoding information about the structural changes moving through the zigzag.

With the right choices of parameters, the 1-dimensional persistence point with the longest lifetime in the zigzag persistence diagram will have birth and death time corresponding to the indices in the zigzag where the Hopf bifurcation appears and disappears.
Lastly, mapping the birth and death times back to the parameter values used to create the corresponding point clouds will give the range of parameter values where the Hopf bifurcation occurs.

Note that there are several parameter choices that need to be selected during the course of the BuZZ method.
First, the dimension $d$ and delay $\tau$ for converting each time series into a point cloud.
Fortunately, there is a vast literature  from the time series analysis literature for this, which leads to standard heuristics.
The second and more difficult parameter is the choice of radius (or radii) for the Rips complexes.
In this paper, the given examples are simple enough that the choice of radii in the BuZZ method can be tuned by the user.
However, in future work, we would like to create new methods and heuristics for choosing these radii.

\subsection{Algorithms}
\label{ssec:algorithm}

While zigzag persistence has been in the literature for a decade, it has not often been used in application, and thus the software that computes it is not well developed.
A C++ package with python wrappers, Dionysus 2\footnote{ \url{https://www.mrzv.org/software/dionysus2/} }, has implemented zigzag persistence; however, it requires significant preprocessing to create the inputs.
We have developed a python package \footnote{\url{https://github.com/sarahtymochko/BuZZ}}
that, provided the collection of point clouds and radii, will perform all the necessary preprocessing to set up the zigzag diagram as shown in Eqn.~\ref{eqn:zz_rips_fixed} to pass as inputs to Dionysus.

\begin{figure}
    \centering
    \includegraphics[width=0.6\textwidth]{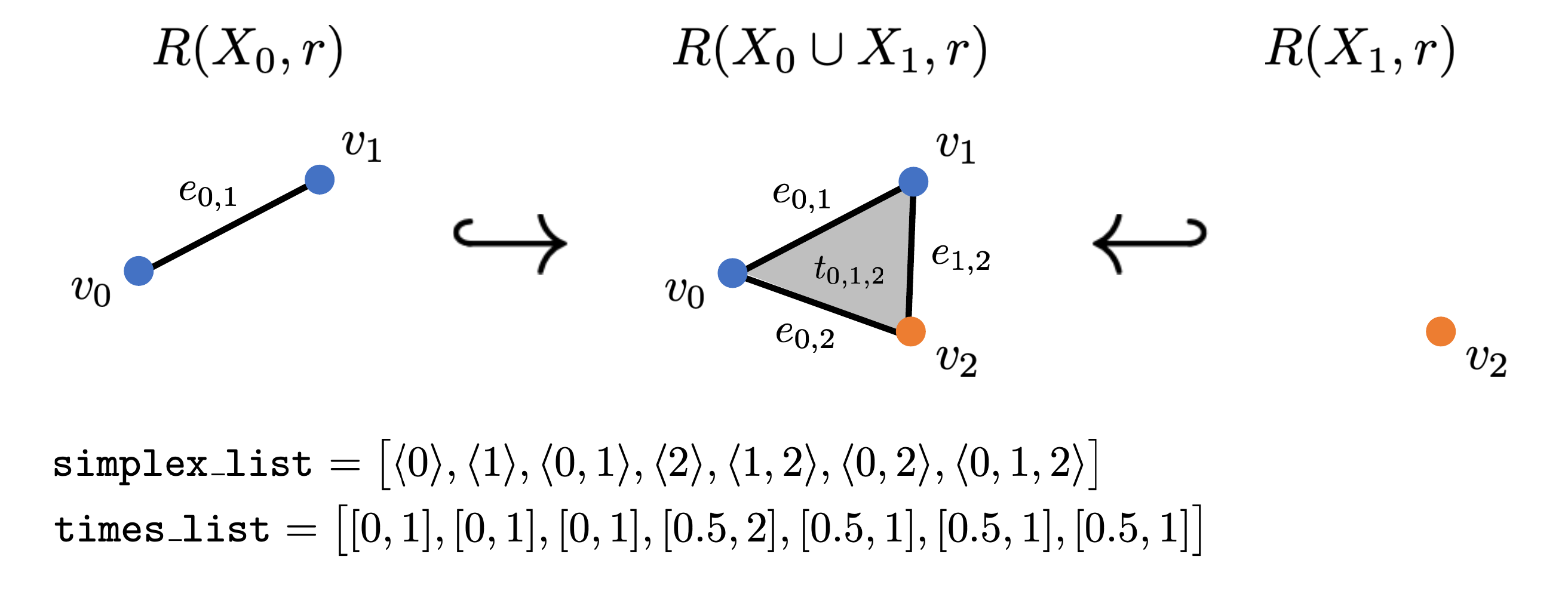}
    \caption{Example zigzag using fixed radius with computed inputs for Dionysus.}
    \label{fig:dio_input_fixed}
\end{figure}

\begin{figure}
    \centering
    \includegraphics[width=0.99\textwidth]{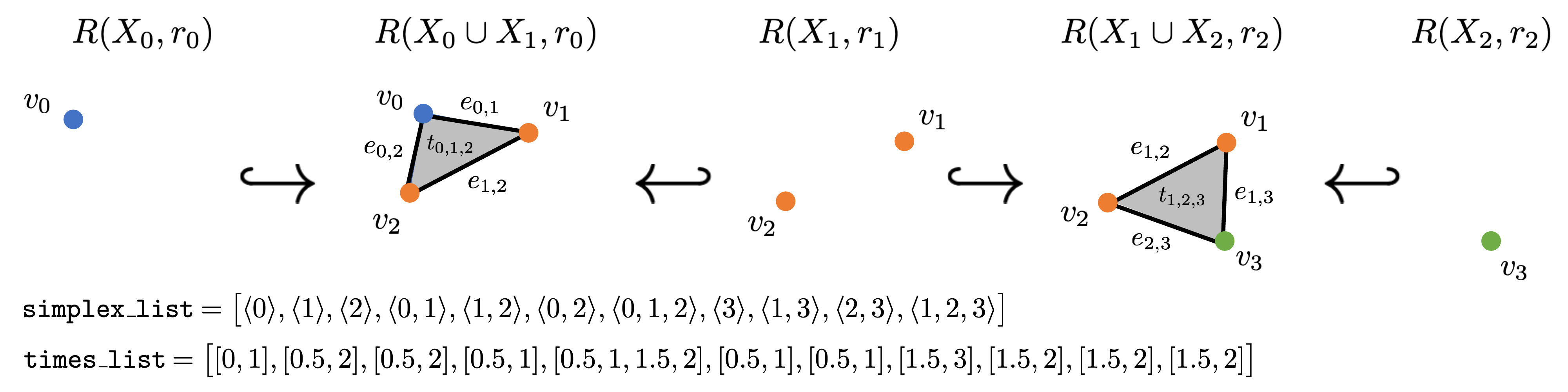}
    \caption{Example zigzag using a changing radius with computed inputs for Dionysus. In this example $r_0>r_1$ and $r_2>r_1$.}
    \label{fig:dio_input_changing}
\end{figure}

Dionysus requires two inputs, a list of simplices, \texttt{simplex\_list}, and a list of lists, \texttt{times\_list}, where the \texttt{times\_list[i]} consists of a list of indices in the zigzag where the simplex, \texttt{simplex\_list[i]}, is added and removed.
A small example is shown in Fig.~\ref{fig:dio_input_fixed}.
Looking at that example, the two vertices and one edge in $R(X_0)$ appear at time 0, and disappear at time 1.
There are two edges and a triangle in $R(X_0\cup X_1)$ that appear there at time 0.5 (recalling that $R(X_i\cup X_{i+1})$ is time $i+0.5$) and disappear at time 1.
Lastly, the one vertex in $R(X_1)$ appears at time 0.5, and never disappears in the zigzag sequence, so by default we set death time to be $2$, which is the next index beyond the end of the zigzag sequence.
This is done to avoid infinite lifetime points, as our zigzag sequences are always finite and an infinite point has no additional meaning.
Note there are other special cases that can occur.
If a simplex is added and removed multiple times, then the corresponding entry in \texttt{times\_list} has more than two entries, where the zero and even entries in the list correspond to when it appears, and the odd entries correspond to when it disappears.
An example with this special case is shown in Fig.~\ref{fig:dio_input_changing} and will be described in more detail later.

If we are using a fixed radius across the whole zigzag, these inputs can be computed rather easily.
In this setting, we only need to compute the Rips complex of the unions, $R(X_i\cup X_{i+1},r)$, which can be done using the Dionysus package, and the list of simplices can be created by combining lists of simplices for all $i$, removing duplicates.
Next, we will outline how to construct the times list.
Starting with the set of simplices in $R(X_i\cup X_{i+1},r)$, we can split them into three groups: (a) simplices for which all 0-dimensional faces are in $X_i$, (b) simplices for which all 0-dimensional faces are in $X_{i+1}$, or (c) simplices for which some 0-dimensional faces are in $X_i$ and some are in $X_{i+1}$.
Because of the construction of the zigzag, all simplices in group (a) appear at time $i-0.5$, since $R(X_i,r)$ also includes backwards into $R(X_{i-1}\cup X_i,r)$, and disappear at time $i+1$, since the union $R(X_i \cup X_{i+1},r)$ is the last time simplices in $X_{i}$ are included.
Similarly, all simplices in group (b) appear at time $i+0.5$, since this is the first time simplices $X_{i+1}$ are included, and disappear at time $i+2$, since $R(X_{i+1},r)$ also includes forward into $R(X_{i+1}\cup X_{i+2},r)$.
Lastly, all simplices in group (c) exist only at $R(X_i\cup X_{i+1},r)$, so they appear at time $i+0.5$ and disappear at $i+1$.
Note that the first case needs to be treated separately, since in $R(X_0\cup X_1,r)$, all vertices in group (a) will appear at $0$.


Using a varied radius, as described in Sec.~\ref{ssec:zigzag}, complicates the above procedure.
Using the same radius, we are guaranteed all simplices in group (a) are in both $R(X_i,r)$ and $R(X_i\cup X_{i+1},r)$, and similarly for group (b), thus we only need to compute $R(X_i\cup X_{i+1}, r)$.
However, with a changing radius this is no longer true.
In the example shown in Fig.~\ref{fig:dio_input_changing}, the edge $e_{1,2}$ appears in both $R(X_0\cup X_1, r_0)$ and $R(X_1\cup X_2, r_2)$ since $r_2>r_1$ and $r_0>r_1$, but it is not in $R(X_1,r_1)$.
Thus, its corresponding list in \texttt{times\_list} is $[0.5,1,1.5,2]$.
Thus the inputs to Dionysus can be computed using the same method as above, except the rips complex needs to be computed for each point cloud, not just the unions, and additional checks need to be done to make sure a simplex being added did not already appear and disappear once before.
If it did, the entry in \texttt{times\_list} needs to be extended to account for the newest appearance and disappearance.

Because of the additional Rips complex computations, and the checks for the special case, the case of a changing radius is significantly more computationally expensive than the case of a fixed radius.
In both cases, there is the computational cost of the zigzag persistence computation as well.
The computational complexity of zigzag persistence is $O(nm^2)$ where $n$ is the number of simplices in the entire filtration and $m$ is the number of simplices in the largest single complex \cite{Carlsson2009}.
Thus, the largest barrier to computation is the zigzag itself, so choosing a radius that is as small as possible without breaking the topology is the goal.

\section{Results}

We will test the BuZZ method on three different examples.
The first example is not based on time series data, but is instead a simple proof-of-concept example to test our methods ability to detect changing circular behavior.
The second example is based on synthetic time series data generated from noisy sine waves of varying amplitude.
This lets us fully utilize the BuZZ method, including the time delay embedding, as well as test resiliency to noise.
The last example is detecting a Hopf bifurcation in the Sel'kov model of glycolysis \cite{Selkov1968}.

\subsection{Synthetic Point Cloud Example}

\begin{figure}[]
    \centering
    \includegraphics[width=\textwidth]{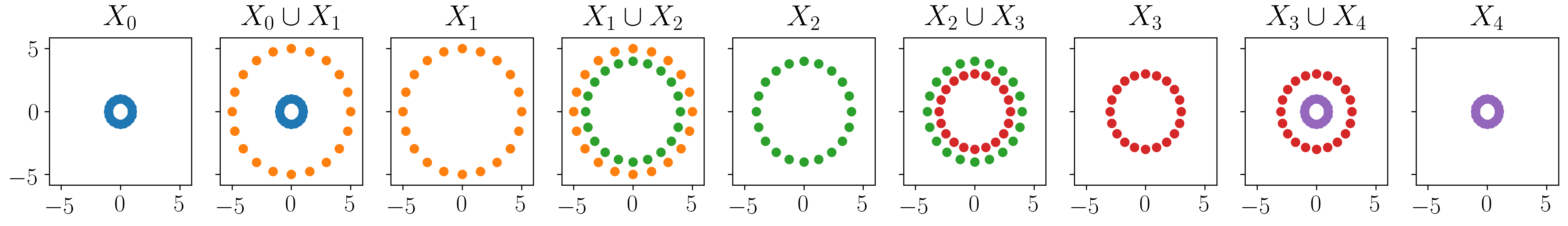}
    \includegraphics[width=\textwidth]{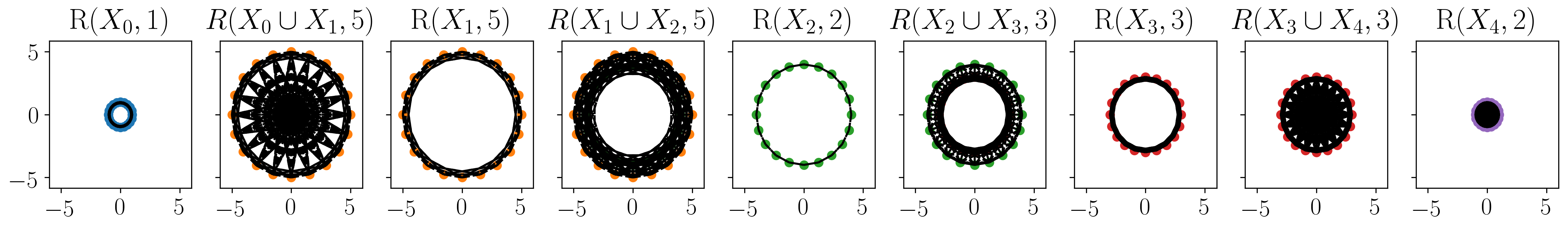}
    \includegraphics[width=0.25\textwidth]{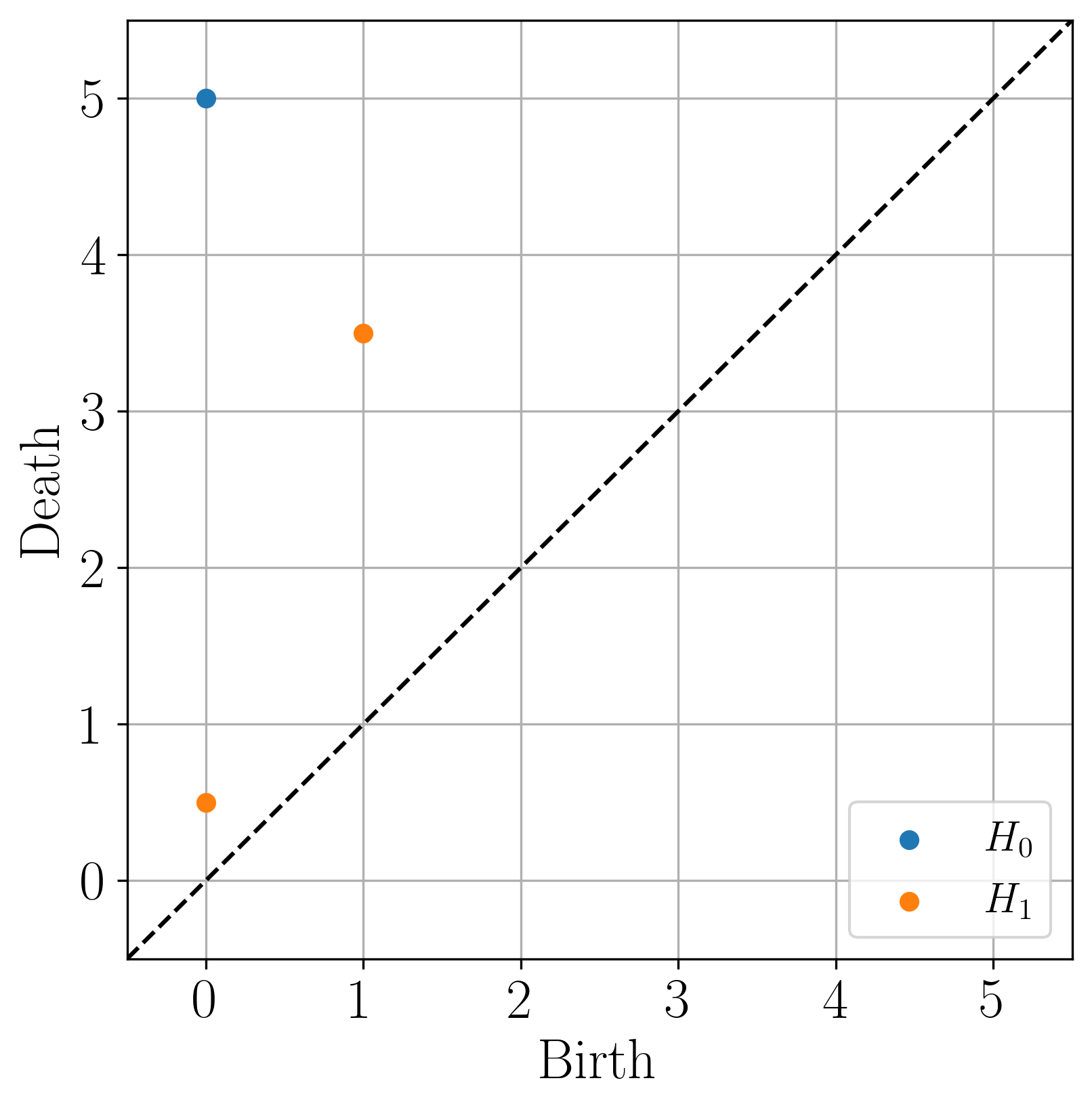}
    \caption{
    Top: Example zigzag of point clouds with unions considered in Sec.~\ref{ssec:SyntheticExample}.
    Middle: Zigzag filtration applied to point clouds using the Rips complex with specified radii. Note that 2-simplices are not shown in the complexes.
    Bottom: The resulting zigzag persistence diagram.}
    \label{fig:circles}
\end{figure}

To start, we will consider a small, synthetic example generating point cloud circles of varying size as shown in  Fig.~\ref{fig:circles}.
Note, because we are starting with point clouds, we skip the time delay embedding step for this example.
While each point cloud is sampled from a circle, the first and last point clouds consist of relatively small circles.
So the strongest circular structure we can see visually starts with $X_1$ and ends with $X_3$.
This is the range we would like to detect using zigzag persistence.

For this example, we will use the generalized version of the zigzag filtration in (\ref{eqn:zz_rips_fixed}) using a changing radii.
Computing the zigzag persistence gives the persistence diagram shown in Fig.~\ref{fig:circles}.
Recall that birth and death times are assigned based on the location in the zigzag that a feature appears and disappears.
Thus, the one-dimensional point $(1,3.5)$ in the persistence diagram corresponds to a feature that first appears at $R(X_1)$ and last appears in $R(X_3)$
Thus, using the persistence diagram we can detect the appearance and disappearance of the circular feature.

This is clearly an overly simplified situation as each point cloud is sampled from a perfect circle.
Next, we will look at a more realistic example.

\subsection{Synthetic Time Series Example}
\label{ssec:SyntheticExample}

\begin{figure}
    \centering
    \includegraphics[width=\textwidth]{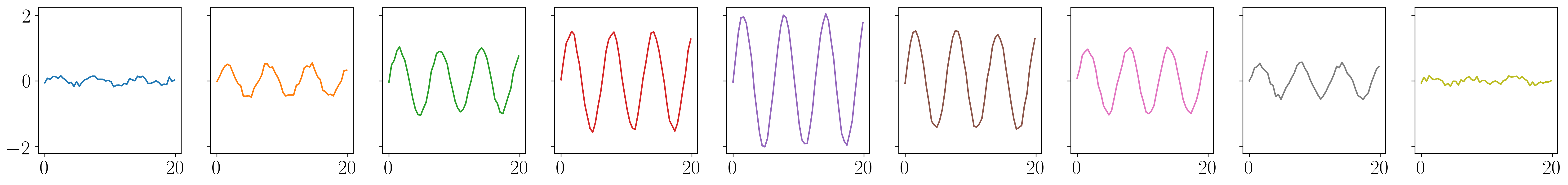}
    \includegraphics[width=\textwidth]{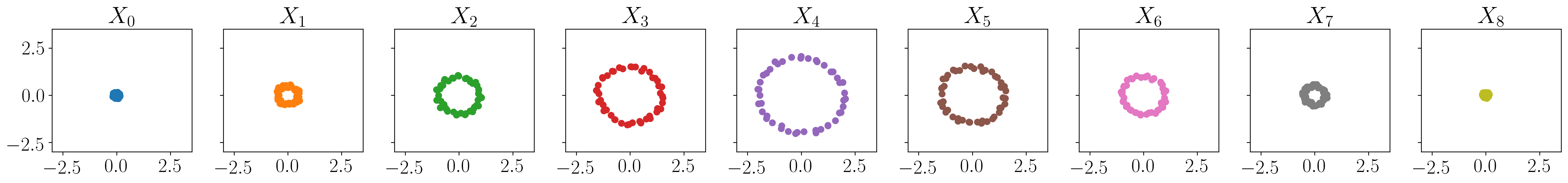}
    \includegraphics[width=0.69\textwidth]{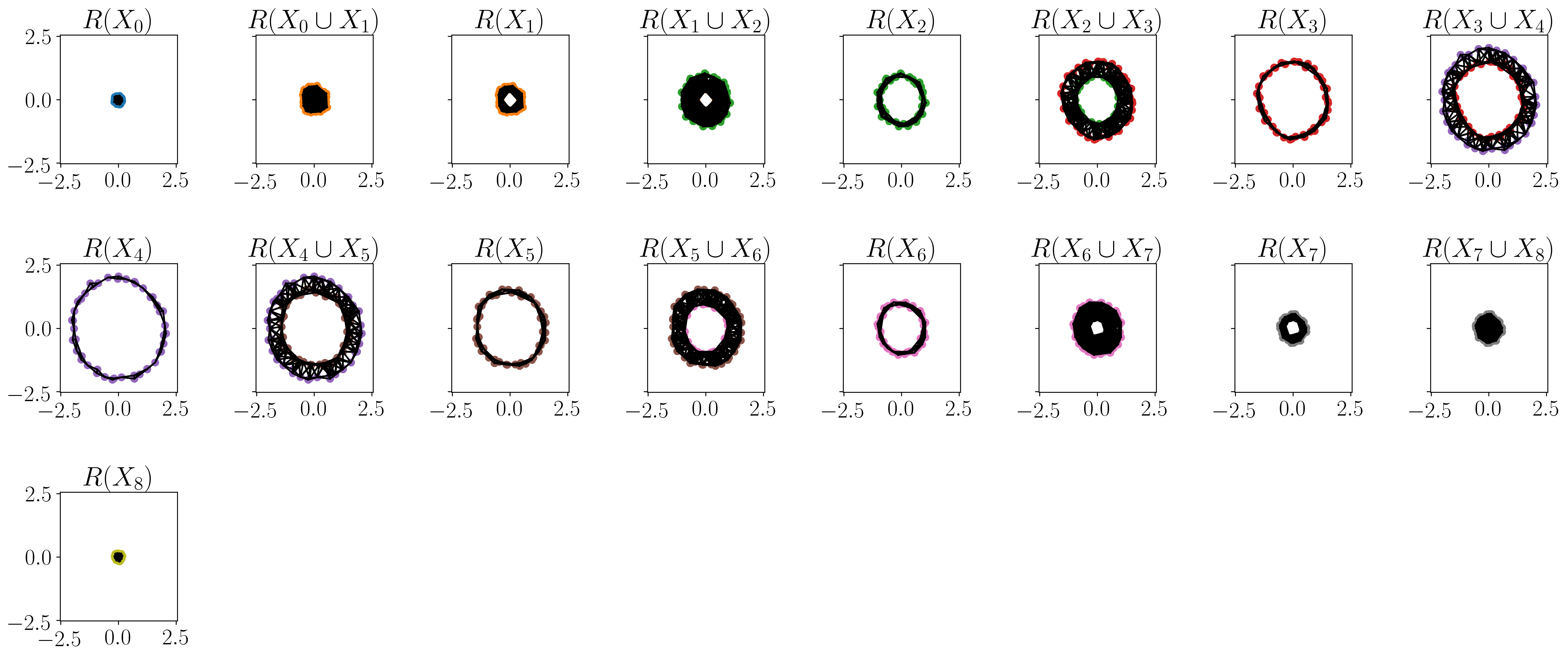}
    \includegraphics[width=0.3\textwidth]{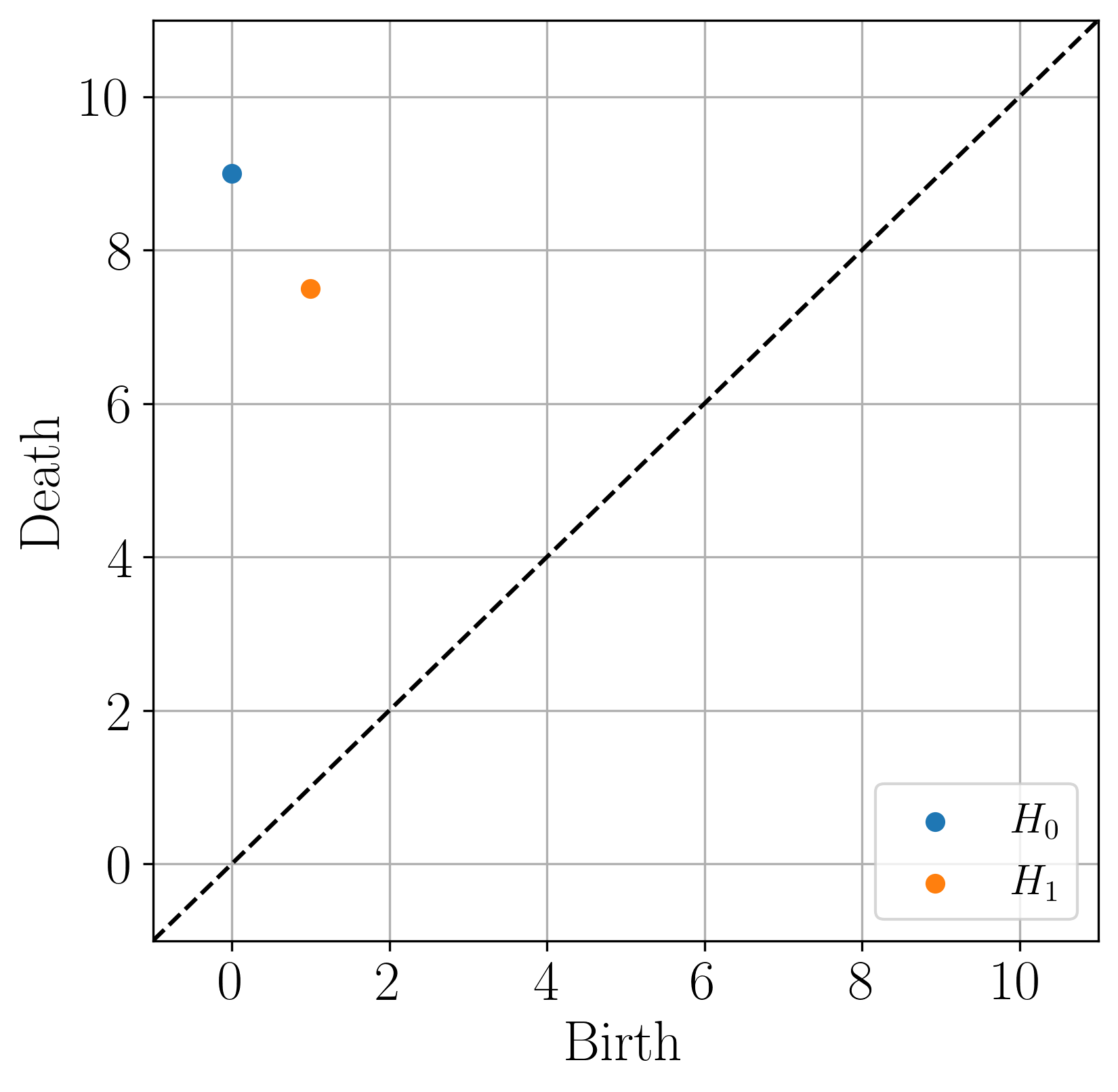}
    \caption{First and second rows: Generated time series data and corresponding time delay embeddings.
    Bottom left: The zigzag filtration using Rips complex with fixed radius of 0.72. Note that 2-simplices are not shown in the complexes.
    Bottom right:the corresponding zigzag persistence diagram.
    }
    \label{fig:sine_ts}
\end{figure}

For the second example, we generate synthetic time series data and apply the full method described in Sec.~\ref{ssec:ourmethod}.
We start by generating sine waves of varying amplitudes and add noise drawn from uniformly from $[-0.1,0.1]$.
The time series are then each embedded using the time delay embedding with dimension $d=2$ and delay $\tau=4$.
The time series and corresponding time delay embeddings are shown in Fig.~\ref{fig:sine_ts}.
Looking at the time series, in the first and last time series any signal is mostly obscured by noise, resulting in a small clustered time delay embedding.
However, for the other time series, the time delay embedding is still circular, picking up the periodic behavior even with the noise.

Next we compute zigzag persistence, resulting in the zigzag of rips complexes and zigzag persistence diagram shown in Fig.~\ref{fig:sine_ts}.
The zigzag persistence diagram has a one-dimensional point with coordinates $(1,7.5)$, indicating the circular feature appears in $R(X_1)$, and disappears going into $R(X_8)$.
This is the region we would expect to see a circular feature.

\subsection{Sel'kov Model}

\begin{figure}[]
    \centering
    \includegraphics[width=0.8\textwidth]{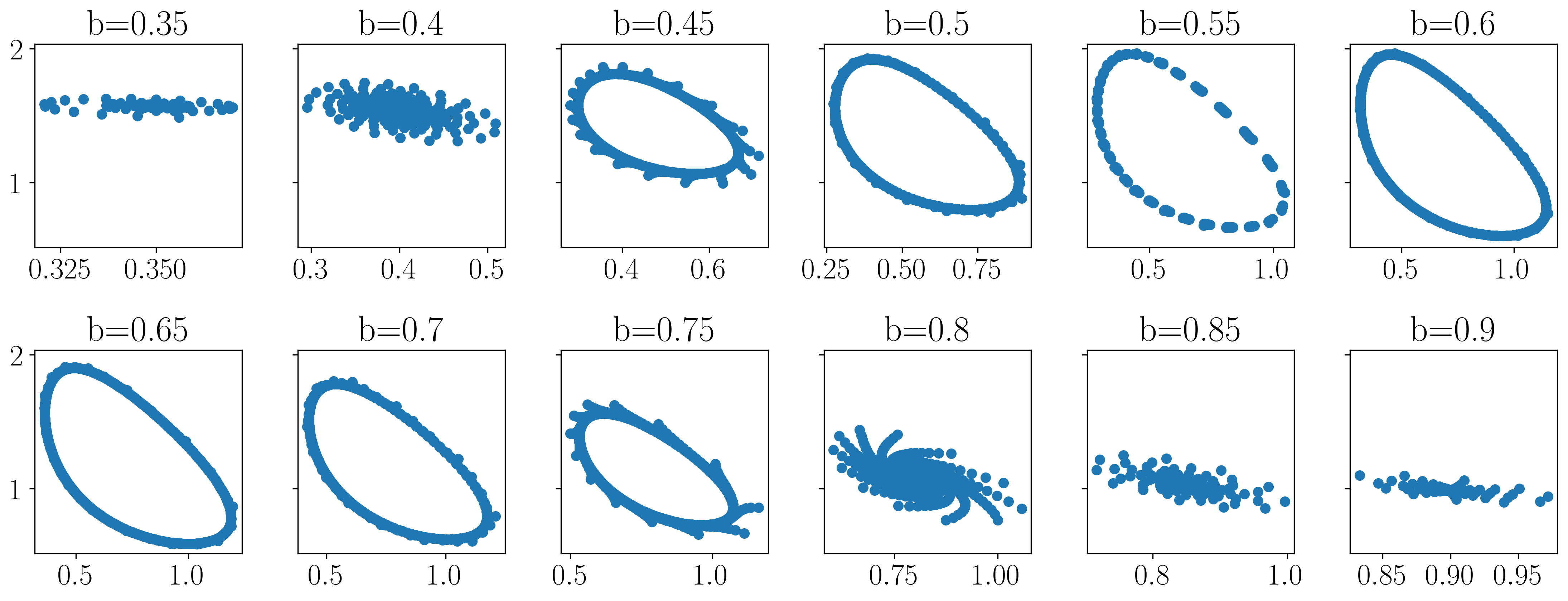}
    \includegraphics[width=0.69\textwidth]{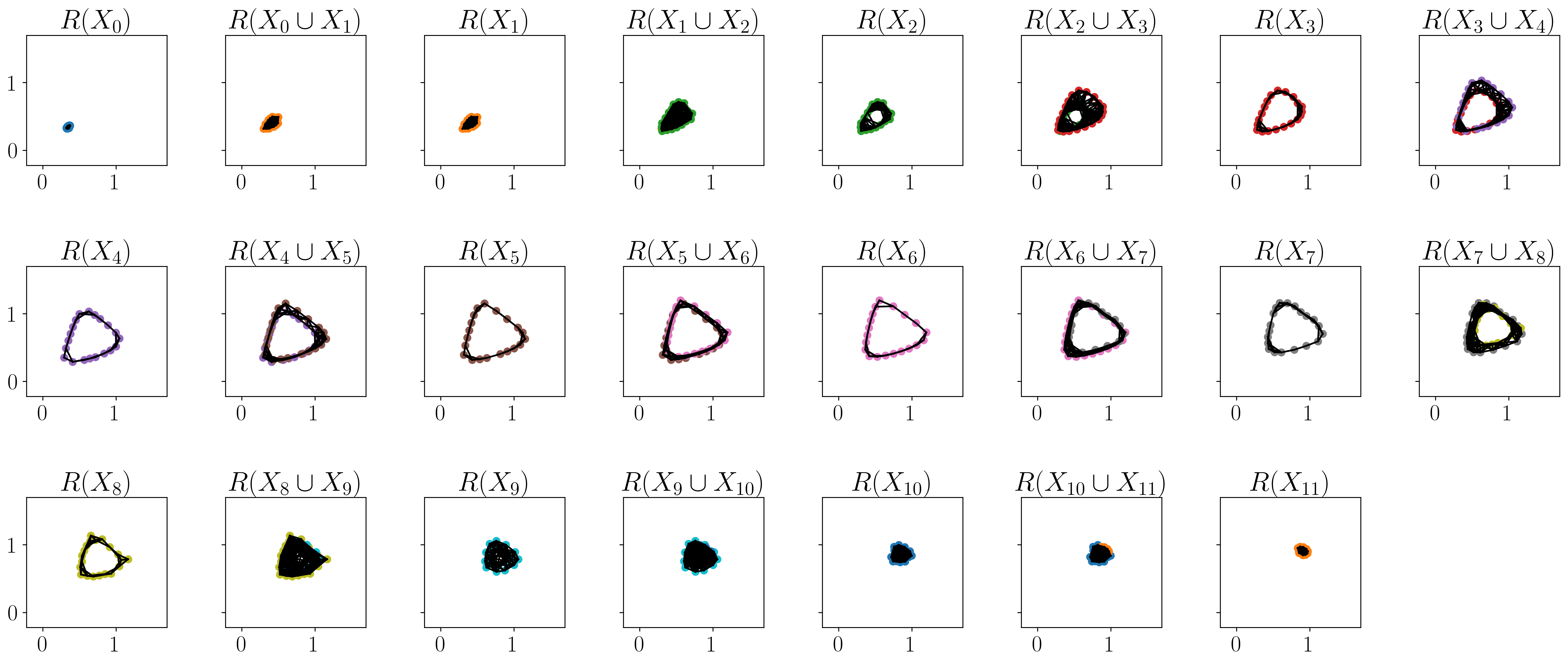}
    \includegraphics[width=0.3\textwidth]{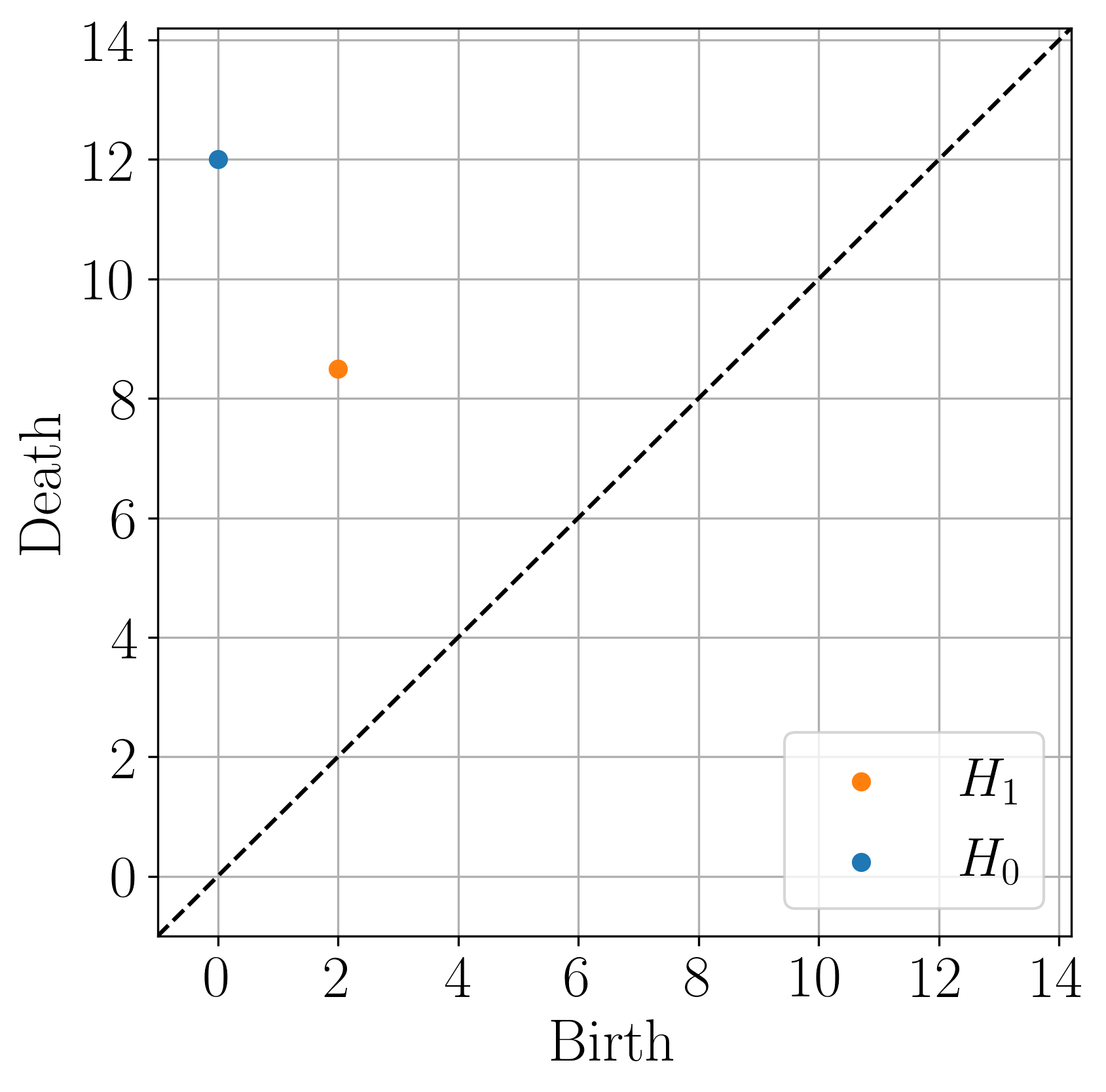}
    \caption{
    Top: Examples of samplings of the state space of the Sel'kov model for varying parameter value $b$.
    Bottom left: zigzag filtration using Rips complex with fixed radius of 0.25. Note that 2-simplices are not shown in the complexes.
    Bottom right: resulting zigzag persistence diagram.}
    \label{fig:selkov}
\end{figure}

Our last experiment is trying to detect a bifurcation in the Sel'kov model \cite{Selkov1968}, a model for glycolysis which is a process of breaking down sugar for energy.
This model is defined by the system of differential equations,
\begin{align*}
    \dot{x} & = -x+ay + x^2y \\
    \dot{y} & = b - ay - x^2y
\end{align*}
where the overdot denotes a derivative with respect to time.
In this system, $x$ and $y$ represent the concentration of ADP (adenosine diphosphate) and F6P (fructose-6-phosphate), respectively.
This system has a Hopf bifurcation for select choices of parameters $a$ and $b$.
This limit cycle behavior corresponds to the oscillatory rise and fall of the chemical compounds through the glycolysis process.

For our experiments, we will fix $a=0.1$ and vary the parameter $b$.
We generate 500 time points of the data ranging between 0 and 500 using \texttt{odeint} in python, with initial conditions $(0,0)$.
We also remove the first 50 points to remove transients at the beginning of the model (this is sometimes referred to as a ``burn-in period'').
This data is constructed using full knowledge of the model, however, in practice, you typically only have one measurement function and then the time-delay embedding is used to reconstruct the underlying system.
To mimic this setup, we will only use the time series corresponding to the $x$-coordinates from the model and use the delay embedding.
These time series are then embedded using the time delay embedding with dimension $d=2$ and delay $\tau=3$.

The next step would be to compute zigzag persistence as described in Sec.~\ref{ssec:zigzag}, however due to the large number of points in the time delay embeddings, this becomes computationally expensive.
In order to reduce the computation time, we subsample these point clouds using the furthest point sampling method (also called a greedy permutation) \cite{Cavanna2015}.
We subsample down to only 20 points in each point cloud, compute the Rips complex zigzag for a fixed radius value of $0.25$, and then compute the zigzag persistence.
Figure~\ref{fig:selkov} shows the zigzag filtration of Rips complexes along with the resulting zigzag persistence diagram.
In the zigzag persistence diagram, the point with the longest lifetime has coordinates $(2,8.5)$.
Again, since these coordinates correspond to the index in the zigzag sequence, this point corresponds to a feature appearing at $R(X_2)$ and disappearing at $R(X_8 \cup X_9)$.
Looking back at which values of $b$ were used to generate these point clouds, we see this corresponds to a feature appearing at $b=0.45$ and disappearing at $b=0.8$.
For the fixed parameter value of $a=0.1$, the Sel'kov model has a limit cycle approximately between the parameter values $0.4 \leq b \leq 0.8$ \cite{Strogatz2014}.
Our method is picking up approximately that same range.

These results use the $x$-coordinates of the model, however the same results can be obtained using the $y$-coordinates and a slightly larger radius value.

\section{Discussion}

Here we have introduced a method of detecting Hopf bifurcations in dynamical systems using zigzag persistent homology called BuZZ.
This method was shown to work on two synthetic examples as well as a more realistic example using the Sel'kov model.
Our method is able to detect the range of the zigzag filtration where circular features appear and disappear.
Thus, this method could be applied to any application with an ordered set of point clouds and a changing topological structure.

While this method has shown success, it also has its limitations.
The method is computationally expensive due to numerous Rips complex computations in addition to the zigzag persistence computation itself.
This issue can be alleviated using subsampling, as shown with the Sel'kov model, but this may not be feasible depending on the application.
Future extensions of this project could include improvements of the algorithms described in Sec.~\ref{ssec:algorithm}.
Additionally, while the method works well in practice, it lacks theoretical guarantees.
Given the method requires parameter choices for the radii of the Vietoris-Rips complexes, we would like some heuristics to be used in practice to choose these radii more easily.
Because our examples in this paper are small, selecting parameters by hand is reasonable.
However, in the future when applied to larger, experimental data, these sorts of heuristics will be necessary.

\section*{Acknowledgements}
The work of EM was supported in part by NSF Grant Nos. NSF CCF-1907591, DMS-1800446.
FAK acknowledges the support of the National Science Foundation under grants CMMI-1759823 and DMS1759824.

\bibliographystyle{plain}
\bibliography{zigzag_bib}

\end{document}